\documentclass[aps,10pt,pra,twocolumn,groupedaddress, floatfix, superscriptaddress,showpacs, showkeys,amsfonts]{revtex4-2}

\usepackage{float}
\makeatletter
\let\newfloat\newfloat@ltx
\makeatother

\usepackage{amsthm}
\usepackage{algpseudocode}
\usepackage{algorithm}
\usepackage{bm}
\usepackage{amsmath}
\newcommand{\be}{\begin{equation}}
\newcommand{\ee}{\end{equation}}

\newcommand{\ket}[1]{|{#1} \rangle}
\newcommand{\bra}[1]{{\langle {#1}|}}

\renewcommand{\phi}{\varphi}
\renewcommand{\Require}{\item \textbf{Input: }}
\renewcommand{\Ensure}{\item \textbf{Output: }}
\usepackage{color}
 \newcounter{definition}

 \newtheorem{Definition}[definition]{Definition}

\usepackage{tabularx}
\usepackage{graphicx}
\usepackage{float}
\usepackage{appendix}

\begin{document}

\title{Discretized Quantum Exhaustive Search for Variational Quantum Algorithms}

\newcommand{\orcidauthorA}{0000-0000-0000-000X} 

\author{$^*$ Ittay Alfassi $^{1}$, Dekel Meirom$^{2}$ and Tal  Mor $^{1,3}$  \\
$^{1}$ Computer Science Department, Technion, Haifa, Israel. \\
$^{2}$ Faculty of Electrical and Computer Engineering, Technion, Haifa, Israel.\\
$^{3}$ The Helen Diller Quantum Center, Technion, Haifa, Israel.\\
$^*$ The authors are ordered in an alphabetical order}


\begin{abstract}

Quantum computers promise 
a great computational advantage over classical computers, yet 
currently available quantum devices have only a limited amount of qubits 
and a high level of noise, 
limiting the size of problems that can be solved accurately with those devices. 
Variational Quantum Algorithms (VQAs) have emerged as a leading strategy to address 
these limitations by optimizing cost functions based on measurement results 
of shallow-depth circuits. 
However, the optimization process usually suffers from severe convergence issues 
as a result of the exponentially large search space --- mainly 
in the form of local minima and barren plateaus.

Here we propose a novel method that can improve variational quantum algorithms
--- ``Discretized Quantum Exhaustive Search''.
On classical computers, exhaustive search, also named brute force, 
solves small-size NP complete and NP hard problems. 
Exhaustive search helps design 
heuristics and exact algorithms for solving larger-size problems 
by finding easy subcases or good approximations.
We adopt this method to the quantum domain, 
by relying on mutually unbiased bases for the $2^n$-dimensional Hilbert space.
We define a discretized quantum exhaustive search that works well 
for small size problems.
We provide an example of an efficient partial discretized quantum exhaustive 
search for larger-size problems, 
for near future and far future goals. 
Our method enables obtaining intuition on NP-complete and NP-hard problems
as well as on Quantum Merlin Arthur (QMA)-complete and QMA-hard problems.

We demonstrate our ideas in many simple cases, providing the energy landscape for 
various problems and presenting two types of energy curves via VQAs.
Using our methods, we learn details that are similar to those obtained by experts' 
knowledge, without using prior knowledge of the problem. 
We expect that merging prior
knowledge and AI/ML tools 
in future work might improve far beyond current abilities of VQA.

\end{abstract}

\maketitle

\section{Introduction}
\label{sect:intro}

In the realm of modern computing, a new paradigm has emerged, promising a great computational advantage over classical computing. Quantum computing, a cutting-edge field at the intersection of physics, computer science, and engineering, harnesses the principles of quantum mechanics to perform computational tasks that are out of reach for classical computers.

The potential applications of quantum computing are vast and include solving complex optimization problems, simulating quantum systems, hacking modern cryptography, and revolutionizing artificial intelligence. However, this exciting promise also comes with substantial challenges, such as managing qubit stability, error correction, and hardware development.

Today's quantum computers are often described as Noisy Intermediate-Scale Quantum (NISQ) computers due to the relatively low numbers of qubits available (e.g. 10's to 100's) and the relatively high levels of noise associated with them (e.g. decoherence and gate fidelity errors) \cite{NISQ, NISQ_challenges_opportunities, NISQ_review}. These limitations result in quantum circuits with short width and shallow depth.  

To make use of these NISQ machines, a class of hybrid quantum-classical algorithms that seek
to leverage the relative strengths of quantum and classical computers together is being developed.
The most common example of such algorithms is Variational Quantum Algorithms (VQAs)\cite{VQA_review}. 
VQAs use a quantum computer to prepare a short-depth Parameterized Quantum
Circuit (PQC) representing a trial solution, also known as an ansatz, to the problem at hand.
Measurements of a final quantum state are used to evaluate a cost function,
which is then minimized on a classical computer to estimate the problem
solution. Prominent examples of VQAs include the Variational Quantum Eigensolver
(VQE) for quantum chemistry and materials applications \cite{VQE_first_paper,
VQE_HEA2017}, Quantum Approximate Optimization Algorithm (QAOA) for
combinatorial optimization problems \cite{QAOA}, and Variational Quantum Linear
Solver (VQLS) for linear algebra problems  \cite{VQLS}. More details about 
these methods are provided in the appendices~\ref{appendix:A},\ref{appendix:C},\ref{appendix:D} and \ref{appendix:E}.

One of the main challenges associated with effective VQA implementations is related to the design of a suitable PQC/ansatz that balances expressibility and noise, while avoiding exponentially vanishing gradients of the cost function, referred to as the barren plateaus problem \cite{Barren_plateaus}.

Designing better ansatzes for VQAs is an active research area that has drawn much attention lately. 
There have been a number of attempts \cite{VQE_best_practices} to improve different ansatz approaches, including ADAPT-VQE \cite{ADAPT-VQE}, Differentiable Quantum Architecture Search \cite{differentiable_search}, Pulse based ansatz \cite{pansatz,ctrl-VQE,ctrl-VQE_multi_levels,PAN,variational_pulse_learning} and Noise-Adaptive Search (QuantumNAS) \cite{QuantumNAS}.

According to \cite{BP_classical_sim}, it might be impossible to avoid barren plateaus using smart ansatz engineering without drastically changing the problem's search space. On the other hand, finding a good initial guess for the optimizer can possibly help avoid barren plateaus. A good initial guess can make the difference between converging to the optimal solution or leading the optimizer to a local minimum trap.
A good initial guess becomes more important as the problem size increases and the problem's Hilbert space and the search spaces become exponentially larger. This increase in search size usually creates exponentially small gradients in the cost function landscape, making the optimization process almost impossible.

For many problems, there are already classical algorithms that can find an approximation to the optimal solution or wide knowledge about the problems at a small scale. These classical approximations, along with expert knowledge, can help create an initial guess for the ansatz for large scale problems.
Still, there are problems where no good classical approximation can be found, and no useful prior knowledge is known about the problem. Therefore, some methods were previously proposed in order to find a good candidate for an initial guess for VQAs without the need to have prior knowledge.

We suggest to use a set of discrete points that are equally spread all over the
search space of the problem. In Section~\ref{subsec:cost_landscape}, we define ``discretized quantum exhaustive search''
(DQES) in analogy to the use of standard exhaustive search in modern computer science.

As far as we know, no previous work has defined DQES 
or used such an approach to learn the cost function landscape when
solving various problems, and as a set of initial states for VQA. 

In addition,
no method to sample the entire Hilbert space using evenly spaced states was defined. 
Using evenly spread states is important for gaining knowledge in a well-organized
way about the landscape of solutions to a problem, and for drawing energy curves
as a function of the optimziation steps, once applying an optimizer for VQA,
especially with the goal of bypassing barren plateaus.

Methods to create good initial states using random states were proposed \cite{parameter_initialization_random, parameter_initialization_random2, parameter_initialization_random3}. Other methods proposed to use machine learning techniques \cite{parameter_initialization_NN, parameter_initialization_baysianlearning, parameter_initialization_transferlearning}. Some ansatzes were proposed to handle barren plateaus instead of avoiding them \cite{ADAPT-VQE, ADAPT_VQE_avoids_BP}, using an adaptive ansatz structure (the ansatz changes during the optimization process, and can escape barren plateaus after entering one).

Another example of an initialization method that is suited for a general VQA is searching over states that can be created with only Clifford gates \cite{Clifford_VQA_1, Clifford_VQA_2}. Using circuits that contain only Clifford gates enables an efficient simulation of the quantum circuits on a classical computer, allowing one to search for an initial guess in a noiseless environment. Clifford states are spread evenly over the Hilbert space, which enables searching in different locations of the problem's space and not only around specific locations. There are far too many Clifford gates in order to use them for a DQES approach (sampling an exponential, but reasonable number of states from the Hilbert space of a problem), as there are $2^{n^2+2n}\prod_{j=1}^{n}(4^j-1)$ Clifford states, where $n$ in the number of qubits \cite{cliffords}. $(2^n)^2$ is much smaller than
$2^{n^2}$, for example, for n=10, $2^{20}$ might be a reasonable number of samples, but $2^{100}$ becomes intractable.

The Clifford states have many benefits, but they are not suitable for every cost function, as the Clifford gates are stabilized by the Pauli operators. The expectation value of a Clifford state against a Pauli operator can lead to either an eigenvalue of $+1$ or $-1$, in case this Clifford is indeed stabilized by that particular Pauli operator, or to $0$. For each Pauli operator, there are exponentially more Clifford states whose expectation value is $0$ compared to those whose expectation value is either $+1$ or $-1$.
Many common VQAs use Pauli expectation values to evaluate their cost function, as it can be implemented very efficiently and easily on a quantum computer.

The main goal of this research is to define novel methods to improve VQA ansatzes and initial states,
leading to better accuracy and faster convergence rates. This goal is achieved by two novel methods:
\begin{itemize}
\item Describing a new method to sample and explore VQAs' problem space, using the DQES method we define later. 
This method allows one to understand the problem's Hilbert space better, creating better guesses for initial parameter choices, and designing methods to avoid barren plateaus and local minima. This method is described in Section \ref{subsec:cost_landscape}. Examples of using this method are presented in Section \ref{sec:results_mub}.
\item Describing an alternative \emph{heuristic} method to explore the VQA's problem space, using an efficient approach. This method aims to bring a heuristic equivalent of the above method for larger problem instances while remaining computationally feasible. This method is presented in Section \ref{subsec:partial_dqes}. Examples of using this method are also presented in Section \ref{sec:results_mub}.
\end{itemize}

\subsection{Mutually Unbiased Bases}
Mutually unbiased bases (MUBs) are orthonormal bases in a Hilbert space that provide complementary and non-redundant information about a quantum state if  measured \cite{first_MUB_paper, MUBs_state_determination, MUBs_count_and_creation_method, MUBs_overview}.
When a pure state is prepared in one basis, a measurement in any other basis of the MUBs provide no information about the prepared state.

\begin{Definition}
For a finite-dimensional Hilbert space of dimension $d$, two bases $\Psi = \left\{\ket{\psi _1}, \ket{\psi _2}, \dots \ket{\psi _d}\right\}$ and $\Phi = \left\{\ket{\phi _1}, \ket{\phi _2}, \dots , \ket{\phi _d}\right\}$ are considered \emph{mutually unbiased} if the bases are orthonormal and the absolute value of the inner product between any vector from the first basis and any vector from the second basis is equal to $\frac{1}{\sqrt{d}}$.
\begin{equation}
    | \langle \psi _i | \phi _j \rangle | = \frac{1}{\sqrt{d}}, \textrm{ for every }i,j \in \{1,2\dots d\}
\end{equation}
\end{Definition}

\begin{Definition}
A set $\{B_1, B_2, \dots , B_m\}$ of orthonormal bases is called a \emph{set of mutually unbiased bases} (a set of MUBs) if each pair of bases $B_i$ and $B_j$ is mutually unbiased.
\end{Definition}

It has been proven that for a Hilbert space with dimension $N$, the size of the largest possible set of MUBs is $N+1$. We call a set of $N+1$ MUBs a \emph{full} set. For any $N$ which is a prime power ($p^n$, for a prime number $p$ and a natural number $n$), it has been shown that a set with such size exists \cite{MUBs_count_and_creation_method, MUBs_state_determination, construction_of_MUBs, MUBs_overview}.

MUBs have applications in various areas of quantum information theory, which use the special properties of the mutual information between the bases.
Some applications use MUBs to maximize the information gained from different measurements, for example in tomography experiments, like state tomography \cite{MUBs_state_determination, MUBs_in_state_determination, MUBs_state_tomography} and process tomography \cite{MUBs_process_tomography}.
Other applications use MUBs to minimize the possible information gained by a third party during quantum communication in order to improve quantum key distribution and quantum cryptography \cite{MUBs_for_qkd, MUBs_for_qkd2}. This is done by using the property of MUBs that the outcome is random when a measurement is made in a basis unbiased to that in which the sent state was prepared. When two remote parties share two non-orthogonal quantum states, attempts by an eavesdropper to distinguish between these by measurements will affect the system and this can be detected.

\section{Methods}
\label{sect:methods}

VQAs search for the optimal solution of the given problem using the search space imposed by the chosen ansatz. The Hilbert space of the problem is determined only by the problem definition and encoding and is not subject to changes by replacing the ansatz. The dimension of this Hilbert space is exponential in the number of qubits required to encode the problem into a quantum computer.
Finding the optimal solution, or even a good approximation to it in such a huge space is usually a very hard task. In many search spaces, the gradients of the cost function become exponentially small in most areas of the search space, a phenomenon called barren plateaus, and most of the states that have significant gradients are concentrated in a small area \cite{Barren_plateaus, BP_and_narrow_gorges}. The concentration of most of the cost function values that have a significant difference between them in a small area makes the search hard even for gradient-free optimizers \cite{BP_in_gradient_free_algorithms, gradient_free_equivalence}.

Some ansatz designs try to narrow this problem by using an adaptive search scheme and by using prior knowledge of the problem to create a relatively small search space inside the exponential problem space \cite{ADAPT_VQE_avoids_BP}. Such prior knowledge is not always available, and even in the cases it does - reducing the search space usually comes with a cost of deeper quantum circuits, which results in more noise.

In our research, we propose methods to learn and gain insights about the values and structure of the cost function in the problem's Hilbert space, in a way that is agnostic to the ansatz that will be used to try to solve the problem. Using such methods can help researchers design ansatzes, optimization algorithms, and initial guesses that can avoid barren plateaus.

The core of these methods is to evaluate the cost function using predefined states that sample the Hilbert space to gain significant knowledge about the cost function landscape.

\subsection{Discretized Quantum Spanning Search}
\label{subsec:DQSS}

Exhaustive search, also named 'brute force', can solve small-size instances of
NP-complete and NP-hard problems. 
By running exhaustive search on many small samples of a problem, as part of the research process, you can try and build heuristics on the general problem.

Here we define discretized quantum spanning search (DQSS) which is useful for small size problems.


The cost function is usually defined using parameterized states $\ket{\psi (\vec{\theta})}$, produced by the ansatz.
In the general case, such a search can be ansatz agnostic.
When disregarding the ansatz, we replace the parameterized states with a chosen set of states from the cost function's domain.
\begin{Definition}
    \textbf{Discretized Quantum Spanning Search (DQSS).}
Choose a set of states $\left\{\ket{\psi _1}\, \ket{\psi _2}, \dots ,\ket{\psi _k}\right\}$ that span the problem's Hilbert space.

Evaluate the cost function of the algorithm $C(\ket{\psi _i})$ for $i \in \{ 1\ldots k \}$.
\end{Definition}

Note that in order to span the space, $k$ must be greater or equal to the dimension of the problem's Hilbert space.
For values of $k$ larger than the dimension of the problem's Hilbert space, the spanning is considered over-complete.

NP problems can be solved in polynomial time by a classical computer once 
a relevant and correct hint is given from an external source. 
In a similar way, the QMA complexity class contains problems that can be solved 
by a quantum computer in polynomial time once a relevant hint is given 
by an external source. 

As explained in appendix~\ref{appendix:C}, in 
VQE, a clever scientist chooses an ansatz and guesses cleverly or randomly an initial state, from which the evaluation and optimization start. 
We notice that VQE resembles QMA, only instead of using the correct hint (as in QMA), a  process of guessing and optimizing is used. 
When a classical or quantum computer tries to solve an NP problem
of a small size, it is possible to test all possible classical hints via exhaustive search, gain knowledge on the structure of the
problem, and design heurstics for larger size problems.
 
As one cannot try all states of the continuous quantum space as possible hints,
we look for an approximated discretized method
that can resemble the classical exhaustive search. 

Searching an exponential number of different states, 
sampled from the problem's relevant Hilbert space, 
for small size problems, may help design heuristics that can then be used for
larger-size problems. Specifically, it may help in studying the energy landscape and energy curves
of problems, hence the locations of barren plateaus and local minima.

\subsection{MUBs for Exploring The Cost Function Landscape}
\label{subsec:cost_landscape}

The states of all of the bases in a full set of MUBs, which we call ``MUB states'', have unique properties that make them good
for extracting information for purposes such as state tomography, 
and process tomography.
In a similar manner, we suggest that these states can be used to 
extract information about the landscape of a cost function of a given problem,
if used appropriately.
A full set of MUB states can be a good candidate as states for a
DQSS for small-size problems. 
They are an exponential number of states that are spread all over the Hilbert
space, and the information learned from them is, in some sense, mutually exclusive.

By measuring the cost function value of a given problem at all these states, one
can get \emph{evenly-spread} (throughout the Hilbert space) information about the structure of the cost function. 
Such information can be properties of states that give similar values (hinting
that there could be barren plateaus areas in the cost function landscape), 
or properties of states that have a larger variety of results (pointing to small areas that might be interesting to search for the optimal result).
It is important to note that such a search is done directly on the cost function landscape, therefore it can be agnostic to the ansatz used.

We define a specific method to do DQSS, by relying on a specific overcomplete spanning that uses all MUB-states of a chosen full MUB set. We call our specific method Discretized Quantum Exhaustive Search (DQES).
\begin{Definition}
\textbf{Discretized Quantum Exhaustive Search (DQES).}
For a problem encoded with $n$ $p$-dimensional qudits with $p$ being a prime number,
choose a full MUB set over the $p^n$-dimensional Hilbert space $\bm{B} = \{B_1, \ldots, B_{P^n+1}\}$.

Evaluate the cost function of the algorithm $C(\ket{\psi^j_i})$ for each state $i$ in each base $B_j$ which is part of the chosen MUB set, $i \in \{1\ldots P^n\}, j \in \{1\ldots P^n + 1\}$.
\end{Definition}

For the case of $n$ qubits, so that there are $(2^n + 1) \times 2^n$ MUB states,
we present an example of DQES in Algorithm~\ref{alg:dqes}, where $j$ runs from zero to $2^n + 1$ and $i$ runs from zero to $2^n$.

\begin{algorithm} \caption{Discrete Quantum Exhaustive Search (DQES) for $n$ qubits}
\label{alg:dqes}
    \begin{algorithmic}[1]
    \Require A cost function $C(\ket{\psi})$ for a problem that can be defined over a $2^n$-dimensional Hilbert space, and a full MUB set $\bm{B} = \{B_1, \ldots, B_{2^n+1}\}$
    \State Define an $n$-qubit register $R$.
    \ForAll {$B_j \in \bm{B}$}
        \ForAll {$\ket{\psi^j_i} \in B_j$}
            \State Set $R$'s state to be $\ket{\Psi} = \ket{\psi^j_i}$.
            \State Calculate $E_{j,i} = C(\ket{\Psi})$.
            \State Save the result $(j, i, E_{j,i})$.
        \EndFor
    \EndFor
    \Ensure the collection of all $(j, i, E_{j,i})$ tuples.
    \end{algorithmic}
\end{algorithm}

In this research we stick to cases in which three MUBs are fully built from product states
(it is possible for $n$ qubits, for any $n \geq 1$ \cite{MUBs_math_construction}).

As the task of evaluating the ground state energy of a Hamiltonian (named the $k$-local Hamiltonian Problem) is Quantum Merlin Arthur (QMA)-complete~\cite{qma,KKR}
and finding the ground state energy of Ising model Hamiltonians is NP-hard~\cite{Lucas2014},
our method enables obtaining intuition on NP-hard problems as well as on QMA-complete 
and QMA-hard problems.

We present an example of the procedure of the DQES method for learning the energy landscape of a VQE problem in Algorithm~\ref{alg:dqesVQE}.
In this example, the cost function is the expectation value of the ansatz state over the Hamiltonian defining the problem.
The results of the method provide an energy landscape of the problem.
One then chooses all or some of the output tuples as the initial point of a variational optimization process of the VQE,
in order to obtain energy curves.

\begin{algorithm} \caption{Discrete Quantum Exhaustive Search (DQES) for VQE}
\label{alg:dqesVQE}
    \begin{algorithmic}[1]
    \Require a Hamiltonian $\mathbb{H}$ defined over an $n$-qubit register, and a full MUB set $\bm{B} = \{B_1, \ldots, B_{2^n+1}\}$
    \State Define an $n$-qubit register $R$.
    \ForAll {$B_j \in \bm{B}$}
        \ForAll {$\ket{\psi^j_k} \in B_j$}
            \State Set $R$'s state to be $\ket{\Psi} = \ket{\psi^j_k}$.
            \State Calculate $E_{j,k} = \bra{\Psi} H \ket{\Psi}$.
            \State Save the result $(j, k, E_{j,k})$.
        \EndFor
    \EndFor
    \Ensure the collection of all $(j, k, E_{j,k})$ tuples.
    \end{algorithmic}
\end{algorithm}

\subsection{Circuit Representation of MUB States}
\label{subsec:mub_circ_representation}
The computation space of a quantum device consisting of $n$ qubits is of dimension $2^n$. It has been shown \cite{MUBs_count_and_creation_method, MUBs_state_determination, construction_of_MUBs, MUBs_overview} that for such a space there exists a set of $2^n + 1$ orthonormal bases which are mutually unbiased. 
Finding such a set of mutually unbiased bases for any number of qubits is not a trivial task. 
Few methods have been proposed to find a full set of unbiased bases, containing $2^n + 1$ orthonormal bases \cite{MUBs_math_construction, MUBs_creation_methods, MUBs_count_and_creation_method, MUBs_state_determination}. 
When using those methods, 
the computational cost of finding orthonormal 
bases which are part of a full MUB set is exponential in the number of qubits. 
For small values of $n$, we noticed one research that found 
efficient ways to implement these MUBs in the lab using adjusted measurement 
equipment \cite{MUBs_physical_implementation}.
In order to perform DQES, one needs to have a quantum circuit that prepares each state in those bases. 
Known algorithms to find a quantum circuit that creates an arbitrary given 
quantum state take exponential time to compute in the worst case 
and might produce a very deep quantum circuit, 
therefore usually cannot be used in the context of VQA.

\begin{figure}[htp]
    \centering
    \includegraphics[width=0.45\textwidth]{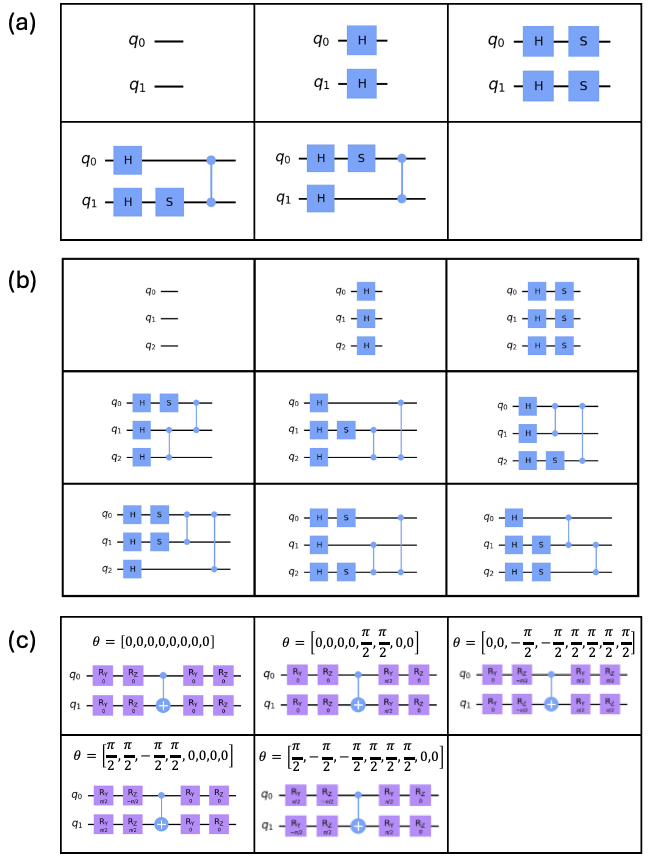}
    \caption{Circuits transforming states from the computational basis to each basis in (a) a 2 qubit MUB set. (b) 3 qubit MUB set. (c) 2 qubit MUB set and the relevant parameters of a hardware efficient ansatz to create these circuits.}
    \label{fig:MUBs_circuits}
\end{figure}

Here we investigate two different circuit representations of MUB states, 
both sufficiently shallow for our demonstrations.
In the first representation, each basis in the MUB set is defined by a 
quantum circuit that transforms a state in the computational basis to a state 
in that basis. 
We have manually found such shallow circuits for each basis in a full set of MUBs 
of 2 and 3 qubits, see 
Fig.~\ref{fig:MUBs_circuits}.

This method provides only a
discrete map, yet it may be used for finding an optimal initial state (or a set
of states), that can then be used as the initial state(s) 
of a parameterized ansatz
controlled by an optimizer as in all usages of VQA.

In contrast with the former method, the second representation relies directly on a chosen ansatz circuit. 
In this representation, each MUB state is defined by the parameter vector 
of an ansatz which prepares the state, hence each MUB state can then directly 
be used by  
utilizing the optimizer to modify the parameters.
When using this method, the chosen ansatz must be expressive enough to allow 
the preparation of each of the states of the MUBs.
In Fig.~\ref{fig:MUBs_circuits} (c), we give a hardware-efficient ansatz circuit for generating 2-qubit MUBs. When acting on the computational state $\ket{i}$, it gives the i'th element of the MUB basis.
Note that the circuit in~\ref{fig:MUBs_circuits} (c) can be used to generate
all MUB 
states \emph{directly} by adding a single layer of NOT gates (where needed)
at the start of the circuit, if the initial state is always the all-zeros
state.

\subsection{Efficient Partial Discretized Quantum Exhaustive Search}
\label{subsec:partial_dqes}

Another tool we suggest and demonstrate via simple examples is an  
efficient partial DQES. Such a search may also provide heuristics 
and potentially lead to exact algorithms for solving larger-size problems, 
by finding easy subcases or better than classical approximations.

The DQES method searches an exponential (in problem size) number of states. 
In this subsection we suggest an alternative method that relies on a polynomial number of states.  
To obtain a scalable method we let some qubits be much more dominant than others 
in the cost function of a VQA problem, yet we do not assume that 
the dominance of each qubit is known a priori. 
We suggest a method to find possible initial states by using a specific set of 
partial-MUB states.

A set of Partial-MUB states is defined as taking a MUB set over $K \ll n$ qubits, encoding the state of the MUB set
into all possible allocations of $K$ indexes in an $n$-qubit register, and keeping all other qubits in some trivial state (WLOG $\ket{0}$).

For small values of $K$ we may generate all such MUB states.
For the scalability proof we shall not assume that $K$ is small.  

For a constant $K$, The number of different options to choose the qubits splitting is 
${{n}\choose{K}} = O\left(n^K\right)$. The number of MUB states for $K$ qubits
is $(2^K+1) \cdot 2^K = (2^K)^2 + 2^K$, leaving us with a total number of
\begin{equation}
    {{n}\choose{K}} \cdot ((2^K)^2 + 2^K) = O(n^K \cdot 4^K) = O(n^K)
\end{equation}

states.

When choosing $K = \log(n)$, the number of MUB state options becomes $n^2+n$,
and the number of different qubit splitting choices is ${{n}\choose{\log(n)}}=n^{\log(n)+O(\log(\log(n)))}$.
The result is super-polynomial but still sub-exponential.

Choosing a state that is in one of the mutually unbiased bases chosen as the trivial state for the $n-K$ qubits that are in a tensor product with the $K$ qubits encoded in a MUB state,  will result in a duplication of states for different qubit allocations.
To avoid such duplication, one can choose a trivial state that is not included in the set of states spanning the chosen unbiased bases.
One can use a-priori knowledge of the problem to choose this static state or to use a random state instead, which can be a different state for each qubit allocation or each basis. One can also calculate the energy for multiple different static or random states for each MUB state and each qubit allocation. Calculating the energy for each qubit allocation and MUB state for a constant number of different states instead of just one does not change the complexity of the calculation, therefore the calculation will still be sub-exponential.

\subsection{MUB States as an Initial Guess for Ansatzes and Optimizers}

One possible outcome from the sampling of the cost function can be a good
initial guess or a set of such initial guesses to use for the optimization process.
These guesses might help avoid barren plateaus and/or local minima and increase the chances of converging into the global minimum. 
The choice for the state to use as an initial guess can simply be the state that gave the minimum value in the cost function evaluation (as long as no additional prior knowledge is known about the problem or the structure of the optimal state), or the best state among a basis whose states showed large differences between their cost values, hinting for an area within the cost function landscape with high gradients.

To the best of our knowledge, the existing gate-based ansatzes' circuits (more details about gate-based ansatzes in appendix~\ref{appendix:B}) do not parameterize all the quantum gates inside them, and all problem-agnostic ansatzes do not have a set of parameters that set the circuit to be equivalent to the identity. As a result, if a circuit that prepares a desired initial guess is inserted at the start of the ansatz, the first sampling point of the optimization will be different from the desired initial guess, even when the zero parameters vector is used.
In order to overcome this, we suggest two methods:
\begin{itemize}
\item \underline{Expressive ansatz} - Use an expressive ansatz that can generate (given the relevant parameters vector)
        each of the MUB states that were used for the energy landscape sampling.
        Then, when wishing to perform an optimization algorithm and use a MUB state $|\psi \rangle$ as an initial guess,
        use the parameter vector that makes the ansatz generate $|\psi \rangle$ as the initial guess for the optimization
         Examples of using such expressive ansatzes to represent MUB states were presented in section \ref{subsec:mub_circ_representation}.
\item \underline{Shifted MUBs} - Choose an ansatz before sampling the energy landscape, let the ansatz be $U(\vec{\theta})$.
        Choose a fixed parameters vector $\vec{\theta}_0$ (for example, the all-zero parameters vector).
        Applying the unitary operation $U(\vec{\theta}_0)$ to a set of MUB states preserves their mutually unbiased properties
        (states from each basis will remain orthogonal, and the inner product of states from different bases will remain the same). 
        Therefore, instead of sampling the energy landscape using MUB states $\ket{\psi_1}, \ket{\psi_2}, \ldots$,
        one can use the shifted MUB states $U(\vec{\theta}_0)\ket{\psi_1},\; U(\vec{\theta}_0)\ket{\psi_2}, \ldots$.
        Then, in order to use $U(\vec{\theta}_0)\ket{\psi_k}$ as an initial guess for the optimization,
        one can use the circuit preparing the MUB state $\ket{\psi_k}$ followed by the ansatz with the fixed parameters vector $\vec{\theta}_0$.
\end{itemize}

As many research papers 
show~\cite{parameter_initialization_random, parameter_initialization_NN, parameter_initialization_transferlearning, parameter_initialization_baysianlearning}, 
choosing a good initial guess (or guesses) for the optimization process of 
a VQA might dramatically affect the convergence of the algorithm.
Namely, it might affect either (or  
both) the number of iterations required for convergence and the quality of the
final result 
to which the algorithm converges.

For some problems, there are domain-specific initial guesses which use prior knowledge of the problem.
For example, a common initial guess for finding the ground state energies of different molecules via VQE is the Hartree-Fock state (note that finding it is an NP-hard problem). Such initial guesses do not always produce the best results, and are only available for problems in which a prior knowledge of the problem is present.

In this subsection, we suggested a method to search for an initial guess that can be
applied to \emph{any} VQA, without the need for prior knowledge, which is useful for
small-size problems. 
For larger problem sizes, by using partial MUB states, 
one can sample a polynomial number of states
in order to find a state which has a better chance to lead to a convergence to an optimal result.
Of course, for larger problems sizes, methods based on prior knowledge or AI/ML tools
can also be highly useful.


\section{Results}
\label{sec:results_mub}

We used the method of DQES on various problems.
For problems that require up to 3 qubits to encode the cost function, we used all the states of a full MUB set.
Calculating the cost function at all of the states of a full MUB set allows us to sample evenly from the entire problem's landscape.
As the number of states in a full MUB set grows exponentially in the number of qubits,
for problems that require more than 3 qubits to encode the cost function we used the partial DQES method,
by using either 2-qubit MUB states or 3-qubit MUB states and keeping the rest of the qubits of the problem in the $|0\rangle$ state.
For each MUB state, we have calculated the cost function analytically without the need to measure multiple copies of the state
(therefore, the result does not have statistical errors).
Such analytical calculation is possible when one has an explicit and efficient representation of the quantum state
and the number of non-identity terms in the operators of the cost function is small.
When using partial DQES, 
the set of states we used for the calculation is the combination of each $K$-qubit MUB state with each possible allocation
of $K$ qubits in the n-qubit register. For example, when using a 3 qubits MUB state for a problem with a size of 4 qubits,
we repeated the process 4 times per each 3-qubit MUB state: each time, another qubit was kept in the $|0\rangle$ state,
while the 3 other qubits were assigned as the MUB state.

The order of the bases in the plots in this section is the same as listed in Chapter \ref{subsec:mub_circ_representation},
in Figure~\ref{fig:MUBs_circuits} (a) and (b).
The order of the states inside each basis is by the computational basis states - $|00\rangle $, $|01\rangle $, $|10\rangle $, $|11\rangle $
for the 2 qubits MUB states and $|000\rangle $, $|001\rangle $, $|010\rangle $, $|011\rangle $, $|100\rangle $, $|101\rangle $, $|110\rangle $, $|111\rangle $
for the 3 qubits MUB states.
Each state in the other bases can be defined as the state resulting from evolving one of the computational basis states
using the basis transformation matrix (or the quantum circuit representing the basis).
The order of the states is defined by the states that are produced from the states in the computational basis in their regular order,
transformed to the other basis in this manner.

\subsection{Single qubit example}
\label{subsec:1qubit-example}

In order to provide intuition regarding the use of MUB-states in VQA, we make
use of the simplest case --- the case of a single qubit.  In this extremely
simple case, we can provide a geometrical interpretation
of the DQES search and the optimization process that
follows it. Unfortunately, we are not aware of a possibility to provide such 
a geometrical intuition for larger dimentsions.

For a single qubit, the canonical MUBs are the $X$, $Y$, and $Z$
bases. Of course, these could be transformed via an arbitrary rotation to any
other, non-canonical, shifted MUB. 
   
Among all VQAs, not many can be applicable to a single qubit. However,
calculating an eigenvector is applicable:
For our demonstration, we choose the Hermitian matrix $\sigma_x + \sigma_y$,
and the goal is to find its eigenstate. The ansatz has three parameters since an
arbitrary single-qubit rotation is defined by the three Euler angles.

\begin{figure*}[t]
\centering
\includegraphics[width=0.8\textwidth]{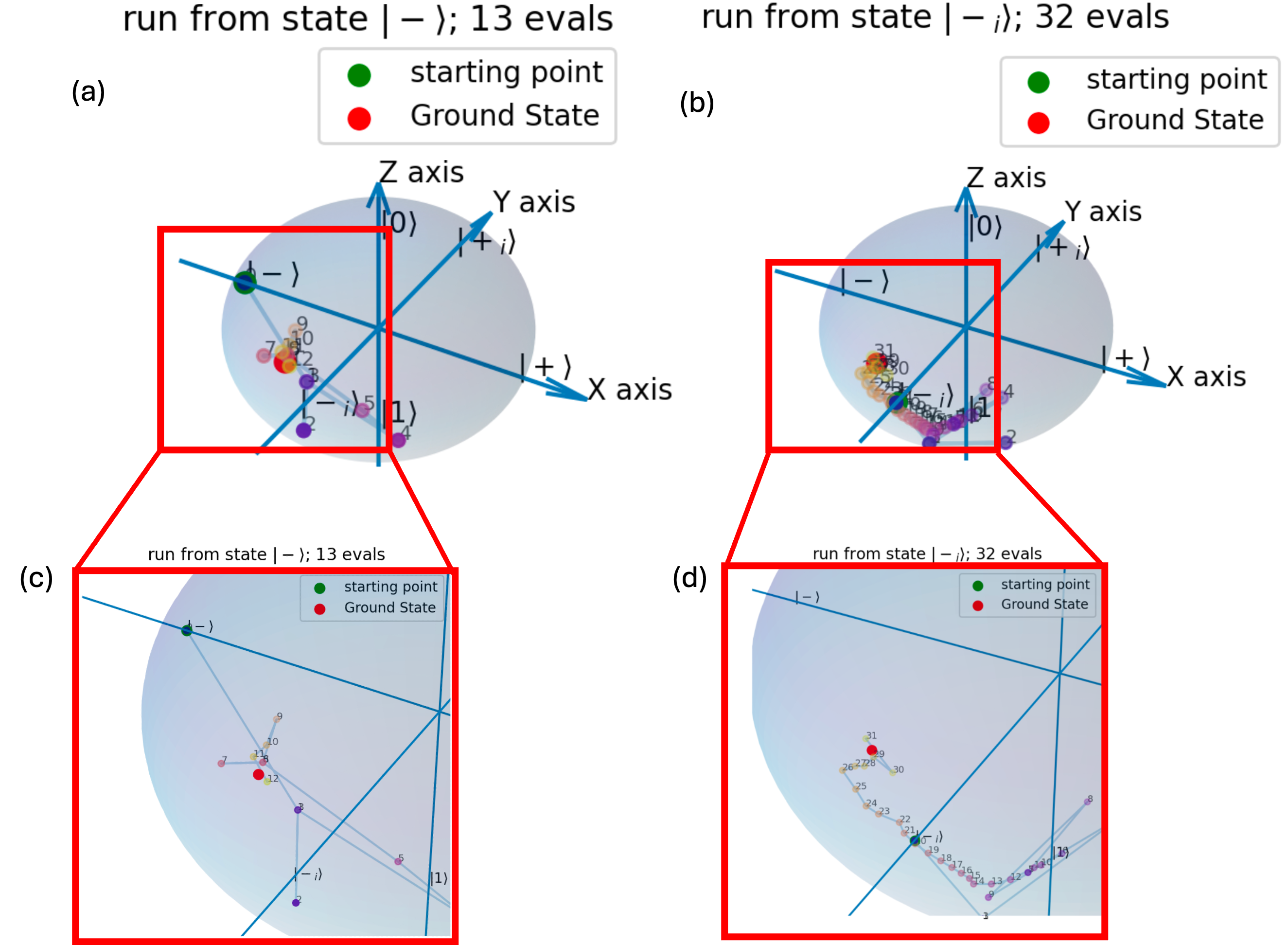}
\caption{Optimization paths (energy curves) 
for 1 qubit search space example. (a) Optimization path obtained by COBYLA 
for a single-qubit example
of the use of MUB states for VQA: the full path, starting from state $|-\rangle $. (b) Optimization path obtained by COBYLA for a single-qubit example
of the use of MUB states for VQA: the full path, starting from the state $|-_i\rangle $. (c) A closeup on reaching the solution for the optimization path presented in (a). 
(d) A closeup on reaching the solution for the optimization path presented in (b).}
\label{fig:1q_visual}
\end{figure*}

We started with each of the six MUB-states, generated using an ansatz,
and we applied the COBYLA optimizer. For each initial state, 
the first guess is the MUB state, and the next three guesses are provided 
(with no optimization attempt) by COBYLA to be a modification of a single angle
of the three parametrized  angles, by a single pre-defined ``distance''. 
Then, starting from iteration number 4, the COBYLA algorithm chooses  
the next guess.  There are two different ways to stop the iterations:
either by the optimizer finding no way to improve the current cost value,
or by the cost function reaching a pre-defined threshold.  
In our examples, the former is used.
We present two examples of reaching the solution ---
starting from two different MUB states, as well 
as a closeup on the path in the Hilbert space (in this case, the Bloch sphere) at the last iterations.
In~figure Fig.~\ref{fig:1q_visual} we see those optimization paths examples. 

\subsection{Molecular electronic structure problems}

We have calculated the energy of various Hamiltonians, and used a full set of
MUB states of 2 qubits for the partial DQES method.
In all of these Hamiltonians we used the Born-Oppenheimer approximation and used the ``qiskit-nature''
Python package~\cite{qiskit_nature} default basis for the orbitals function approximation (STO-3G basis),
solving the full configuration interaction (FCI) problem (for more details about the problem definition, see appendix~\ref{appendix:D}).
For choosing the amount of orbitals that are considered in the calculation
(which varies for the different molecules tested), we also used the package's default for this basis.
We used the default ``method type'' for the driver class of the ``qiskit-nature'' package,
which is a restricted method, that assumes that the energy of the spin up and spin down spin-orbitals
of the same orbital is equal, resulting in these spin-orbitals being occupied or empty together.
The Hamiltonians were converted into spin Hamiltonians using parity
transformation \cite{parity_and_bravyi_kitaev}, and encoded into qubits after
utilizing symmetries in the Hamiltonian to reduce the amount of qubits needed in
order to encode the solution by 2 (as described in \cite{reduce_qubit_count,
VQE_best_practices}). All Hamiltonian coefficient calculations were done
using the ``qiskit-nature'' Python package~\cite{qiskit_nature}, version 0.4.3.
When we have calculated the Hartree-Fock states for comparisons, we used the
same parameters (STO-3G basis as the basis for the orbitals and the same atomic
distance) and used the ``pyscf'' Python package~\cite{pyscf}, version 1.6.3, to
perform the calculation. Some preliminary details are provided in appendix~\ref{appendix:D}.

In the results presented below, we chose Hamiltonians representing the molecules $H_2$, $HeH^+$ and $LiH$.
Given the basis we used (STO-3G) and the ``method type'' (RHF),
the software automatically chooses $k$ lowest-energy orbitals ($k$=2,2,6, for the three molecules respectively) meaning $2k$ spin orbitals.
This means orbitals [0,1] for $H_2$ and $HeH^+$ and orbitals [0,1,2,3,4,5] for $LiH$. 
For $H_2$ and $HeH^+$ there are just two electrons and both are analyzed, in the four spin-orbitals.
For $LiH$, four electrons participated in the calculation.

The first Hamiltonian represents the $H_2$ molecule at an atomic distance of $0.75 [\text{\AA}]$.
The relevant orbitals included in the calculation are the molecular orbitals formed from the 1s orbitals from each $H$ atom,
resulting in 4 spin orbitals.
Using the parity transformation and utilizing symmetries to reduce the number of qubits required in the encoding, we converted this Hamiltonian into a qubit Hamiltonian of 2 qubits, in the tensored Pauli operators form:
\begin{equation}
\begin{split}
& \mathbb{H}_{H_2}^{r=0.75} = -1.05540303 \cdot II + 0.38874759 \cdot IZ \\
& - 0.38874759 \cdot ZI - 0.01117714 \cdot ZZ \\
& + 0.18177154 \cdot XX \\
\end{split}
\end{equation}

\begin{figure}[t]
\centering
\includegraphics[width=0.48\textwidth]{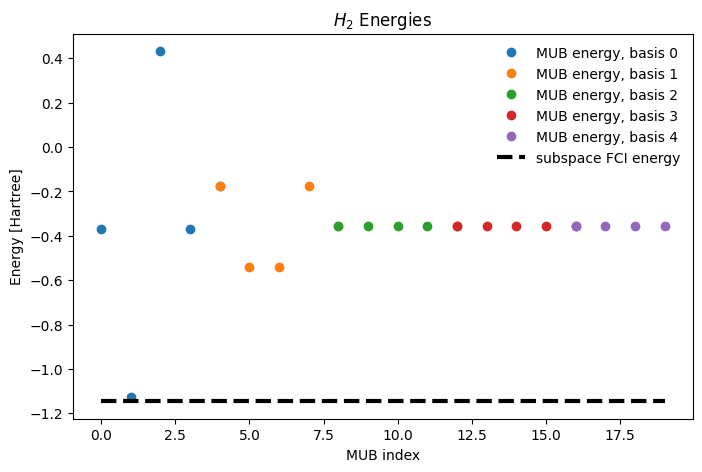}
\caption{Exhaustive search of all the 2 qubit MUB states and their energy with respect to the $H_2$ molecule in $0.75 [\text{\AA}]$ atomic distance.}
\label{fig:H2_MUB}
\end{figure}

\begin{figure*}[t]
\centering
\includegraphics[width=1\textwidth]{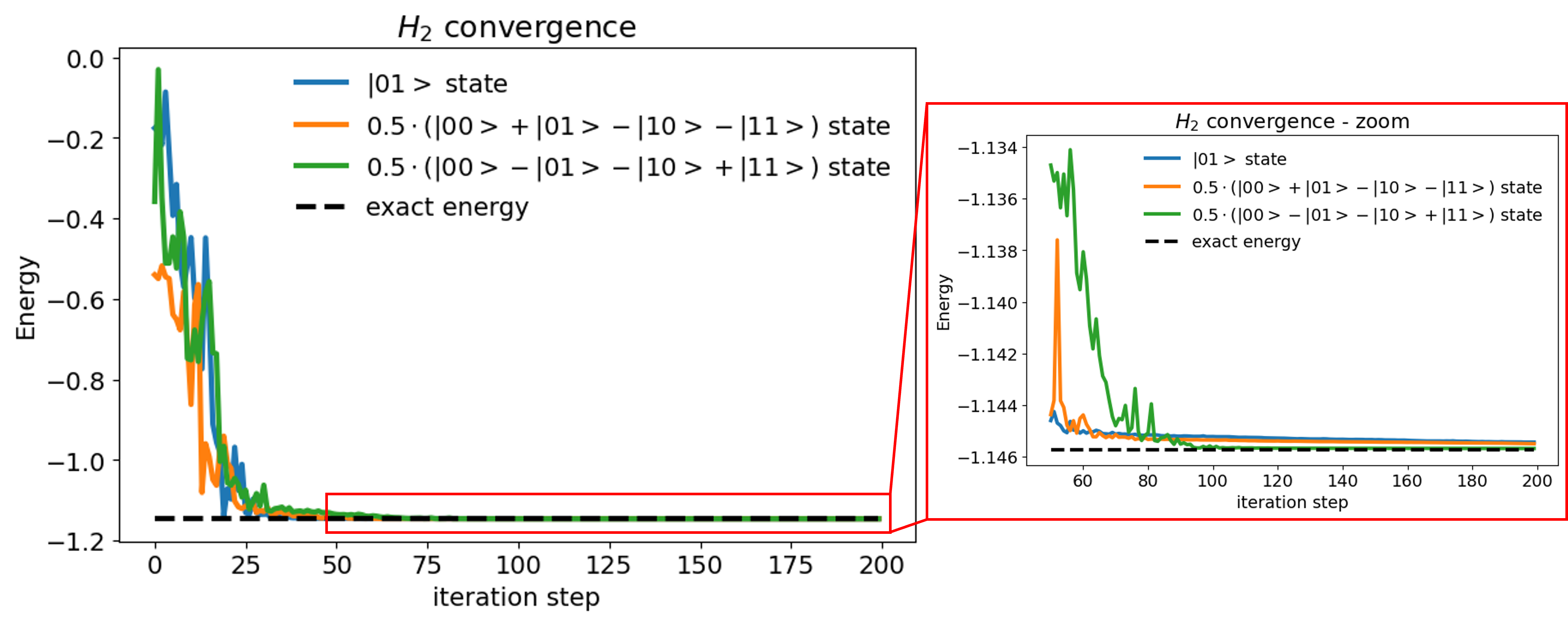}
\caption{Convergence results of the optimization process of a VQE algorithm on the $H_2$ molecule in $0.75 [\text{\AA}]$ atomic distance using a noiseless simulator, the COBYLA optimizer, and different initial states. The initial states are the 3 MUB states that had the lowest energy in the landscape calculation of the molecule. The right figure is a zoom to the final iterations of the optimization.}
\label{fig:H2_convergence}
\end{figure*}

The results are shown in Fig.~\ref{fig:H2_MUB}. The best MUB state result is the same as the Hartree-Fock state energy.
By using the Real Amplitudes Hardware Efficient Ansatz with a single layer,
and using the 3 best MUB result as an initial state, we were able to achieve final results within chemical accuracy
(within $1.6 [mH]$ from the FCI results) for all three states. 
The ansatz is expressive enough to create all of the MUB states,
so we used the parameters form of the MUB states representation to initialize the ansatz with these states,
as shown in Fig.~\ref{fig:MUBs_circuits}~(c).
We achieved these results using a noiseless simulator and the COBYLA optimizer.
The convergence results (energy curves)
are shown in Fig.~\ref{fig:H2_convergence}. All three states converged to approximately the same result.

The second Hamiltonian represents the $HeH^+$ molecule at an atomic distance of $1 [\text{\AA}]$. The relevant orbitals included in the calculation are the molecular orbitals formed from the 1s orbitals from both the $H$ atom and the $He$ atom, resulting in 4 spin orbitals. By utilizing symmetries we managed to reduce the number of qubits required in the encoding to 2.
The resulting Hamiltonian, in the tensored Pauli operators form, is:
\begin{equation}
\begin{split}
& \mathbb{H}_{HeH^+}^{r=1} = -3.04506092 \cdot II  \\
& + 0.50258052 \cdot IZ + 0.11926278 \cdot IX - 0.50258052 \cdot ZI \\
& + 0.11926278 \cdot XI - 0.13894646 \cdot ZZ - 0.11926145 \cdot ZX \\
& + 0.11926145 \cdot XZ + 0.11714671 \cdot XX \\
\end{split}
\end{equation}

\begin{figure}[t]
\centering
\includegraphics[width=0.48\textwidth]{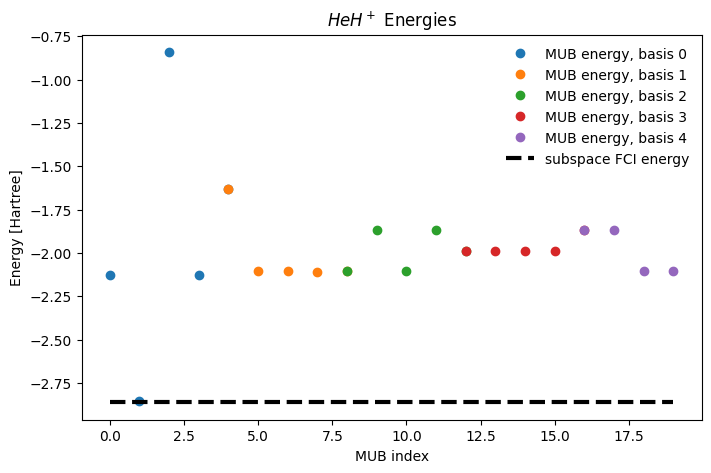}
\caption{Exhaustive search of all the 2 qubit MUB states and their energy with respect to the $HeH^+$ molecule in $1 [\text{\AA}]$ atomic distance.}
\label{fig:HeH_MUB}
\end{figure}

\begin{figure*}[t]
\centering
\includegraphics[width=1\textwidth]{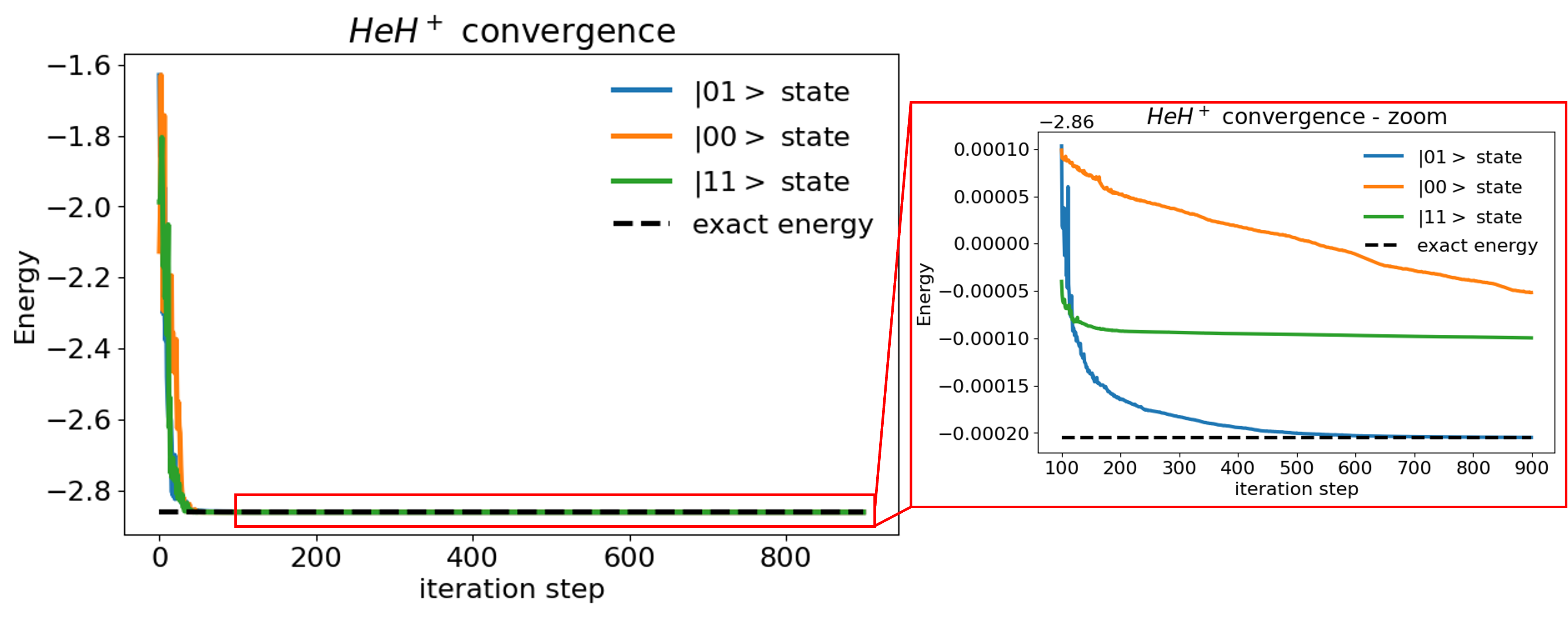}
\caption{The convergence results (energy curves)
of the optimization process of a VQE algorithm on the $HeH^+$ molecule's Hamiltonian in $1 [\text{\AA}]$ atomic distance using a noiseless simulator, the COBYLA optimizer,
and different initial states. The initial states are the 3 MUB states that had the lowest energy in the landscape calculation of the molecule.
The right figure is a zoom to the final iterations of the optimization.}
\label{fig:HeH_convergence}
\end{figure*}

The results are shown in Fig.~\ref{fig:HeH_MUB}. The best MUB state result is the same as the Hartree-Fock state energy.
By using the Real Amplitudes Hardware Efficient Ansatz with a single layer, and using the 3 best MUB result as an initial state,
we were able to achieve final results within chemical accuracy for this problem as well. We achieved these results using the same simulator and optimizer.
The convergence results are shown in Fig.~\ref{fig:HeH_convergence}. Although all three states converged to an energy within chemical accuracy,
there is a slight energy difference between the converged energies. The state that had the lowest energy in the landscape calculation ($|01\rangle$)
converged to the lowest energy in the optimization process.

The third Hamiltonian represents the $LiH$ molecule at an atomic distance of $1.5 [\text{\AA}]$.
10 qubits were required to represent this Hamiltonian when taking all the relevant orbitals (6 orbitals, which represent 12 spin orbitals) into consideration.
We used the partial DQES method, defined in \ref{subsec:partial_dqes}. We used 2-qubit MUB states, and kept the rest of the qubits at the $|0\rangle$ state.
For each experiment, we tested all possible pairs of qubits to be defined as the qubits of the MUB state.
We tested how the results vary when reducing the amount of orbitals which we took into consideration in the encoding.
Each removed orbital reduces the amount of needed qubits in the encoding by 2.
The FCI result shown in the figures resembles the exact ground state energy of the tested subspace - the Hamiltonian formed only by the considered orbitals.
The ground state energy changes when considering different orbitals or different basis functions, as shown in \cite{LiH_energies}.

When analyzing the 12 spin-orbitals case, four electrons participated in the calculation.
To analyze the options with less orbitals,
we used the \texttt{ActiveSpaceTransformer(num\_electrons, num\_spatial\_orbitals, active\_orbitals}) class 
of the ``qiskit-nature'' package.
If one defines \texttt{num\_electrons = 2} without reducing the number of
orbitals and not specifying specific active orbitals (\texttt{active\_orbitals = none}), only two electrons are analyzed while the other
two fill in the lowest level [0], which leads to analysis of orbitals [1,2,3,4,5].
If one define \texttt{num\_orbitals = 2} while also reducing the number of active electrons via \texttt{num\_electrons = 2}, the two active
electrons are analyzed in orbitals [1,2].

\begin{figure*}
\centering
\includegraphics[width=0.95\textwidth]{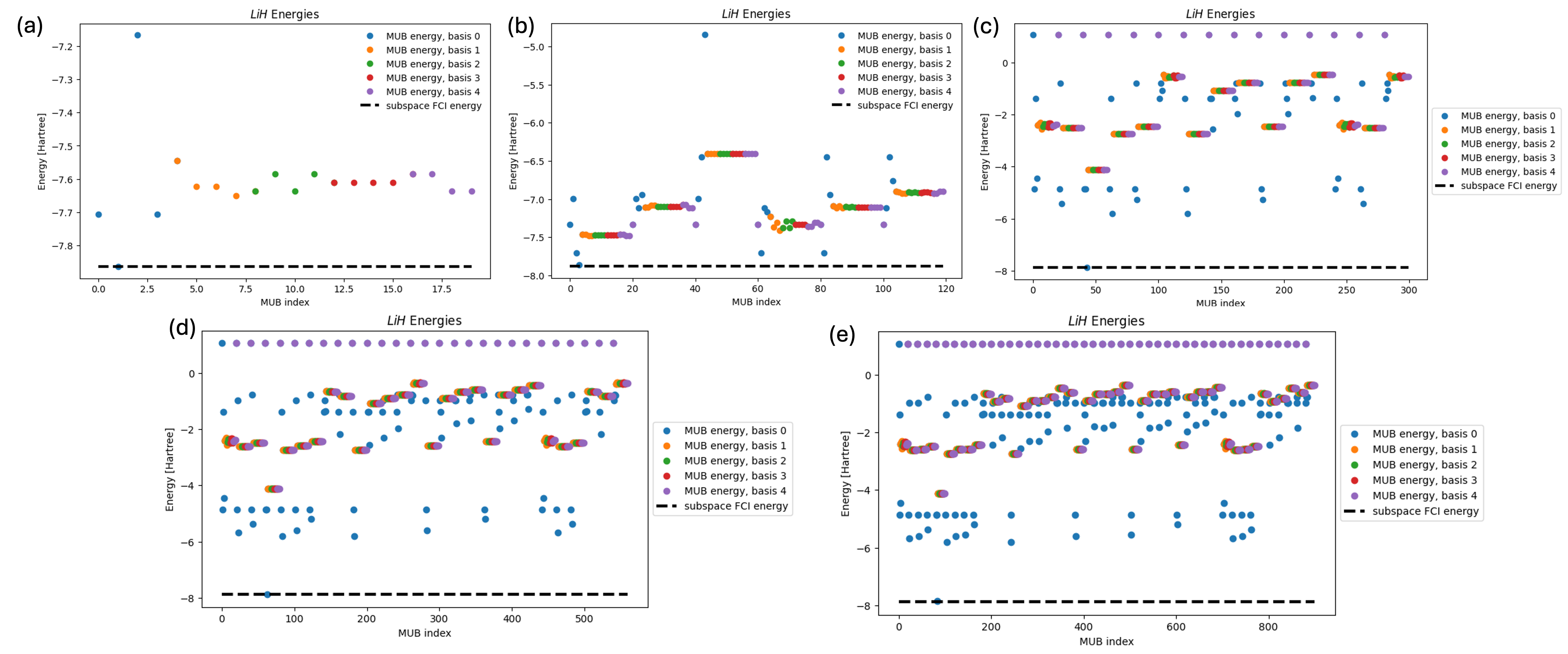}
\caption{Exhaustive search of all the 2 qubit MUB states and their energy with respect to the $LiH$ molecule in $1.5 [\text{\AA}]$ atomic distance when considering different number of orbitals in the calculation. The points from different qubit allocations are showed one after the other, while the ordering of the points for allocation is the same as described as the beginning of section \ref{sec:results_mub}. (a) considering 2 orbitals in the calculation. 2 qubits are required in this encoding. Total of 20 points were calculated. (b) considering 3 orbitals in the calculation. 4 qubits are required in this encoding. 6 qubit allocations were tested, with a total of 120 points calculated. (c) considering 4 orbitals in the calculation. 6 qubits are required in this encoding. 15 qubit allocations were tested, with a total of 300 points calculated. (d) considering 5 orbitals in the calculation. 8 qubits are required in this encoding. 28 qubit allocations were tested, with a total of 560 points calculated. (e) considering 6 orbitals in the calculation. 10 qubits are required in this encoding. 48 qubit allocations were tested, with a total of 900 points calculated.}
\label{fig:LiH_MUB_results}
\end{figure*}

We checked the option with no reduced orbitals, and the option mentioned above, as well as
options with specifically chosen orbitals. Thus, the analysis we did was
for orbitals [0,1,2,3,4,5], [0,1,2,3,5], [0,1,2,5], [1,2,5] and [1,2]. This leads
to using 10, 8, 6, 4, and 2 qubits, respectively, where in each of
those, we saved two qubits relative to a naive parity encoding using
symmetries \cite{reduce_qubit_count,VQE_best_practices}. Results are shown in Fig.~\ref{fig:LiH_MUB_results}.
The best MUB state result is the same as the Hartree-Fock state energy in all of those cases.

From these results, it can be seen that the orbitals whose occupation is the most relevant for the optimal result are orbitals 1 and 2. In addition, it can be seen that the most relevant information is found in the computational basis.
Both of these results are known when using the prior knowledge of quantum chemistry.
Such information might be also gained for problems where no such prior information is found,
like some cases of systems of linear equations, which can be solved using the VQLS algorithm. By gaining such information, an initial guess can be made for the algorithm, and a better choice of ansatz can be made (for example - choosing an ansatz that does not use phase gates at all, like the Real Amplitudes ansatz).

\subsection{Transverse-Field Ising Problem}
We used MUB states to sample the energy landscape of transverse-field Ising Hamiltonians with random coefficients for the $ZZ$ and $X$ terms.
The Transverse-Field Ising model Hamiltonian is 
\begin{equation}
\mathbb{H} = \sum _{i,j} c_{ZZ} Z_i Z_j + c_X X_i
\end{equation}
while $c_{ZZ}$ and $c_X$ are the coefficients of the $ZZ$ and $X$ terms respectively, which were chosen randomly in our tests.

\begin{figure}[t]
    \centering
    \includegraphics[width=0.48\textwidth]{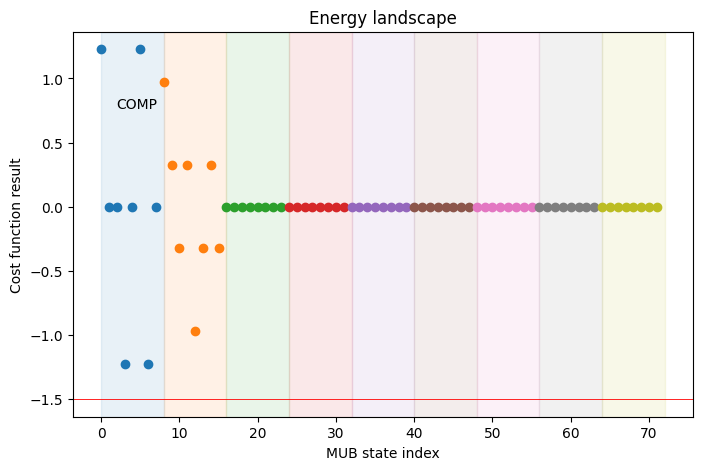}
    \caption{Energy landscape of a transverse-field Ising Hamiltonian. each point represents a different state, while all of the states from the same basis have the same color. The lowest energy state is part of the computational basis (blue color). The Hamiltonian coefficients are: $\mathbb{H}=0.04645122 \cdot ZZI + 0.04645122 \cdot IZZ + 0.27498273 \cdot XII + 0.27498273 \cdot IXI + 0.27498273 \cdot IIX$}
    \label{fig:ising_comp}
\end{figure}

\begin{figure}[t]
    \centering
    \includegraphics[width=0.48\textwidth]{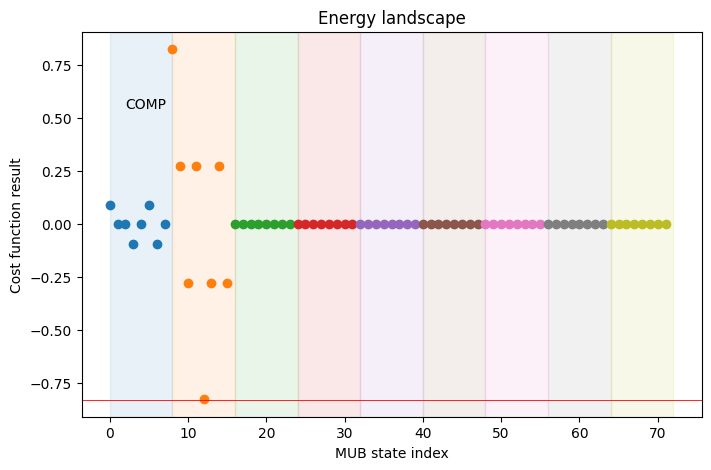}
    \caption{Energy landscape of a transverse-field Ising Hamiltonian. each point represents a different state, while all of the states from the same basis have the same color. The lowest energy state is part of the Hadamard basis (orange color). The Hamiltonian coefficients are: $\mathbb{H}=0.61436456 \cdot ZZI + 0.61436456 \cdot IZZ + 0.32435029 \cdot XII + 0.32435029 \cdot IXI + 0.32435029 \cdot IIX$}
    \label{fig:ising_had}
\end{figure}

In Figures~\ref{fig:ising_comp} and~\ref{fig:ising_had}, we show results of energy landscape calculation
of such Hamiltonians with 3 qubits. The landscape calculation is done using a full MUB set.
We colored the results from different bases and showed that by having different parameters for the Hamiltonian, the basis that has a state with the lowest energy is different.
Because of the locality of each term in the sum of operators that create the Hamiltonian, MUB states from high dimension can be traced out to dimension 2 and their expectation value can be calculated exactly using a classical computer. Such calculation can be done only if the state is known and can be represented directly as a state-vector and not only as a unitary transformation that evolves some initial state (which in that case such transformation should be calculated first, and in case of a very high order, might not be possible to calculate on a classical computer). For that reason, such a calculation can't be used during the optimization process after using a MUB state as the initial state for an ansatz.

\subsection{Combinatorial Problems}

We used the efficient partial DQES method with a full set of MUB states of 3 qubits to explore the landscape of the Max-Cut problem~\cite{Barahona1988} (more details in appendix~\ref{appendix:E}).
We used 3-qubit MUB states, and kept the rest of the qubits at the $|0\rangle$ state. For each experiment, we tested all possible triples of qubit locations to be defined as the qubits of the MUB state.

\begin{figure*}
\centering
\includegraphics[width=0.95\textwidth]{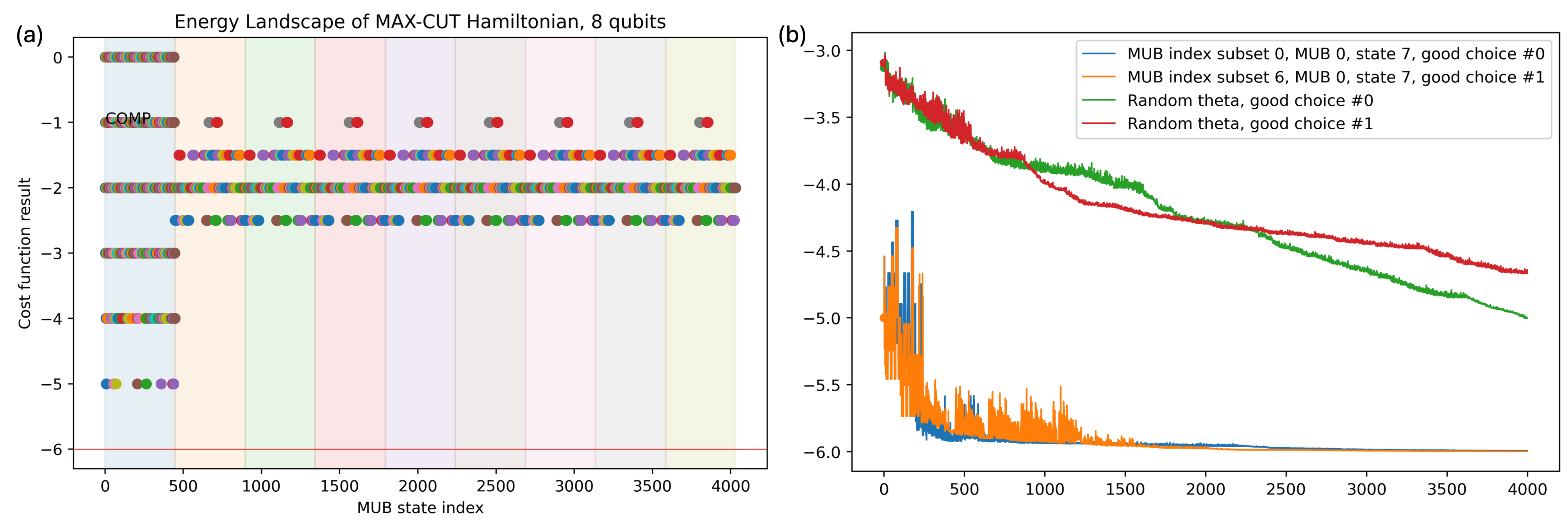}
\caption{Max-Cut problem example. (a) Exhaustive search of all the possible tensor products of 3 qubit MUB states with the rest of the qubits in the $|0\rangle$ state, and their cost function value for a random 8 nodes Max-Cut problem. The points of all the states from each basis of the MUB from all possible qubit allocations share the same background color. The ordering of the points for each MUB is the same as described as the beginning of section \ref{sec:results_mub}. 8 qubits were used to encode this problem and 56 qubit allocations were tested. For each qubit allocation, the energy for 72 different MUB states was calculated, a total of 4032 points calculated. (b) Convergence graphs of VQE runs on the 8 nodes Max-Cut problem, starting with different initial guesses - 2 choices of MUB states and 2 random states. }
\label{fig:maxcut_example}
\end{figure*}

For our example, we used a random graph with 8 nodes, and sampled its energy landscape using partial-MUB states.
We can see from the spread of the different results, that the cost function has the most differences in values in the computational basis.
In the other bases, in contrast, most of the states yield the same result. Such knowledge can be used to limit the ansatz to specific states in the Hilbert space of the problem.
We also tried to find the optimal solution by running an optimization process using the VQE algorithm (using COBYLA as optimization algorithm and hardware efficient ansat with 3 layers and linear entanglement).
We compared the results between different initial states: 2 MUB states which had low energy in the energy landscape calculation step, and 2 random states which had low energy relative to other random states that were randomized (we sampled the same amount of random states and MUB states).
To create the random states we sampled vectors of parameters for the ansatz. For each parameterized gate we randomized a rotation angle uniformly from the range $[0,2\pi)$. For each such random parameters vector we calculated the cost function for the problem's graph. We than ran the VQE optimization process twice, using the parameters from the two vectors which resulted in the lowest cost as initial parameters for the optimization.
The results are shown in Fig.~\ref{fig:maxcut_example}

\section{Conclusion}
\label{sect:conclusion}
In this paper, we tried to tackle local minima and barren plateaus, which are the most dominant problems
of VQAs --- one of the promising algorithmic approaches to the NISQ era of quantum computing. 

We have developed a method to efficiently sample the Hilbert space of the cost
function of a VQA problem, using the ``efficient partial DQES'' approach. We used MUB states in order to extract maxinmal information out from such sampling, in an ansatz-agnostic method. We tested our method on multiple different problems, including different molecular electronic structure problems, Transverse-Field Ising model problems, and Max-Cut problems.

We showed that such a method can be used in order to learn characteristics of the landscape
of the cost function and some useful information about the given problem at hand,
such as good approximations of the problem and good initial guesses to the solution.
We managed to learn good approximations and initial guesses regarding the tested problems, which are similar to those achieved by experts in the problems' fields. We have extracted such information without using prior expert knowledge.
We believe that the information learned on the cost function landscape can be used to help avoid barren plateaus during the optimization process done in a VQA.

\subsection{Open problems for Future research}
Our research introduces new ansatzes and methods to deal with the convergence issues of VQAs, but further research is still needed in order to make these algorithms achieve results better than the best classical computers using today's available quantum hardware.
Such further research might include -
\begin{itemize}
\item Improve DQES and efficient partial DQES and try these 
on additional problem types.
\item In contrast with the trivial choice of the $\ket{0}$ state for non-MUB qubits in efficient partial DQES,
        use a random choice (from a pre-determined set) for the single-qubit states
        and explore the implications for different problem types.
\item Try to use the information gained from the DQES search or the efficient partial DQES to avoid barren plateaus and local minima and improve the algorithm convergence rate or final accuracy.
\item Apply DQES and partial DQES to problems with larger numbers of qubits.
\item Merge our methods with experts' prior knowledge of specific problems.
\item Merge our method with various AI and ML tools. 
\item Check our methods with various optimizers, and define novel types
of optimizers to deal with the set of MUB states or a set of partial MUB states.
\item Further investigate whether the efficient
partial DQES does better than random states for NP problems as seen in two
examples we provided in Fig.~\ref{fig:maxcut_example}.
\item Further investigate advantages and disadvantages of DQES and
efficient partial DQES when compared to random initial states of the
same amount, for problems beyond NP, and for problems in BQP (such as
variational quantum factoring).
\end{itemize}

\paragraph{Acknowledgements:}
D.M. and T.M. thank the Helen Diller Quantum Center at the Technion for their generous Support.
T.M. and I.A. thanks the Quantum Computing Consortium of Israel Innovation
Authority for financial support. 

\bibliographystyle{plain}
\bibliography{mub_cites}


\appendixpage
\appendix

\section{Variational quantum algorithms}
\label{appendix:A}

Variational quantum algorithms represent a pioneering approach within the field of quantum computing, developed to tackle the challenges that arise in the NISQ era and offer a practical bridge between current limited quantum hardware and problem-solving \cite{VQA_review}.
At their core, these algorithms employ quantum circuits with tunable parameters, often referred to as variational circuits or parametric circuits, to address a wide range of computational tasks. By iteratively adjusting these parameters and leveraging classical optimization techniques, variational quantum algorithms seek to find the optimal configuration that minimizes a specific objective function. This hybrid approach combines the power of quantum superposition and entanglement with classical optimization's robustness, making it suitable for tackling problems in fields as diverse as quantum chemistry, optimization, machine learning, and materials science.

The various variational quantum algorithms are defined mainly by their cost function, and the structure of the associated quantum circuit which allows the efficient evaluation of that cost function. The efficiency is measured in the ability to calculate the cost function using only a polynomial (in the problem size) number of circuit executions at each iteration of the classical optimization process.

The evaluation of the cost function is done by calculating the expectation values of some operators
derived from the cost function. These expectation values are calculated in a state
$|\psi (\vec{\theta})\rangle$, which is a parameterized state prepared by the used ansatz.
\begin{equation}
    C(\vec{\theta}) = \sum_k f_k(\langle \psi (\vec{\theta})| O_k |\psi (\vec{\theta})\rangle)
\end{equation}
Where $f_k$ is a set of functions, $O_k$ is the set of the operators that are calculated on the state prepared by the ansatz.

At each iteration, a classical computer calculates the next parameter vector that will be used
in the ansatz parameters. 
The quantum computer repeatedly measures the operators $O_k$ over the parametric state.
A classical computer evaluates the functions $f_k$ and constructs the value of the cost function $C$.
The new evaluation is fed into the optimization algorithm, which outputs the next set of parameters.
This process continues until a convergence or other stop criteria (like a maximum number of iterations)
is reached. The goal of this process is to find an optimal parameters vector:
\begin{equation}
    \vec{\theta}_{opt} = \arg \min _{\vec{\theta}} C(\vec{\theta})
\end{equation}

This iterative process has several benefits that are important in the NISQ era.
Among those benefits is the ability to break down a deep circuit into a batch of shallower circuits,
which in turn allows to calculate the result with less noise. Another important benefit is the partial
resilience to coherent errors that arise from this iterative process \cite{VHQC_theory, VQE_error_robustness}.
This resilience is related mainly to the ansatz part, and is a result of the fact that the prepared state
is changing at each iteration to a state that gaints a better value from the cost function,
fixing coherent errors by using new parameters.
For example, if the computer has an over-rotation error for one of its gates as a result of a miscalibrated gate,
the parameters vector can shift slightly in the following iterations to account for this over-rotation.
VQAs still have many challenges, as achieving the global optimum of ansatz parameters can be computationally intensive, and the landscape of the objective function can feature many local minima. In many cases, as the problem size increases, the gradients of the cost function become exponentially small, making the search for the optimal solution almost impossible. This challenge is referred to as the barren plateaus problem \cite{Barren_plateaus}.

\section{Ansatzes}
\label{appendix:B}

An ansatz is typically constructed as a quantum circuit with a specific structure,
such as a sequence of quantum gates, where some or all of the gates have adjustable parameters.
These parameters are the variables that can be optimized to find the best approximation to the target
quantum state, which in turn optimizes the algorithm's cost function.
The ansatz prepares a parameterized quantum state -
\begin{equation}
    |\psi (\vec{\theta })\rangle = U(\vec{\theta })|\psi _0 \rangle
\end{equation}
while $|\psi (\vec{\theta })\rangle$ is the result parameterized quantum state, $U(\vec{\theta })$ is the parameterized quantum circuit (the ansatz) and $|\psi _0 \rangle$ is some initial quantum state.

The term "ansatz" is derived from the German word for "assumption" or "postulate," and it reflects the idea that you are making an educated guess about the form of the quantum state you want to prepare.

Many VQAs can evaluate their cost function using various ansatzes, each such ansatz having its advantages and disadvantages. The choice of an ansatz is a critical design decision in VQAs. It influences the algorithm's convergence speed, accuracy, and suitability for a given problem. It also determines the quantum circuit's expressiveness - its ability to represent complex quantum states or solutions to optimization problems. Different ansatz structures are suited to different types of problems and may have varying levels of flexibility and computational efficiency.

For many ansatz designs, there is a trade-off between the quantum circuit's expressiveness, the depth of the quantum circuit, and the number of parameters associated with it, which might influence its convergence rate.

In some cases, domain-specific ansatz circuits are designed with prior knowledge about the problem at hand. These ansatzes are usually referred to as problem-inspired ansatzes. For instance, in quantum chemistry applications, one might use an ansatz tailored to represent the electronic structure of a molecule. An example of such a tailored ansatz for the electronic structure of a molecule is the Unitary Coupled Cluster (UCC) ansatz. Problem-inspired ansatzes usually have relatively small search space (only between states that are meaningful to the problem) which allows for fast convergence of the algorithm. On the other hand, in many cases this comes with the disadvantage of a deep quantum circuit, creating a strong noise that makes the results inaccurate.

Another common type of ansatzes is the hardware-efficient ansatz. These ansatzes are generic and can be used in multiple problems and algorithms. The structure of these ansatzes is designed to minimize the resource requirements and circuit depth of the ansatz, while still providing a reasonably expressive representation of quantum states. This structure usually also takes into account the native gates set of the device and avoids the use of complex multi-qubit gates. By that, complex gate decompositions are avoided, which can introduce additional errors and increase the gate depth.
The term "hardware-efficient" underscores the aim of creating quantum circuits that can be implemented efficiently on existing or near-term quantum hardware, which typically has limitations in terms of qubit connectivity, gate fidelities, and gate times.

\section{Variational Quantum Eigensolver (VQE)}
\label{appendix:C}

The Variational Quantum Eigensolver (VQE) \cite{VQE_first_paper, VQE_best_practices} is a variational algorithm whose goal is to find the smallest eigenvalue of a given Hermitian matrix (usually described as a Hamiltonian of a system).
\begin{equation}
    \mathbb{H} |\psi \rangle = E_0 |\psi \rangle
\end{equation}
The algorithm finds $E_0$, which is usually referred to as the ground state energy of the given Hamiltonian.

The algorithm's input is a decomposition of the Hermitian matrix into a polynomial number of measurable observables (usually tensored Pauli operators), a chosen ansatz, and an initial guess for the initialization of the parameters.
The input Hamiltonian is represented as -
\begin{equation}
    \mathbb{H} = \sum_{i}^{L} h_i O_i
\end{equation}
where $L$ is a polynomial number in the number of qubits $L=poly(n)$, $O_i$ are measurable observables, each with a coefficient $h_i$ (each observable can act on all or some of the qubits). 

The algorithm uses the variational principle, which states that the expectation value of a Hamiltonian, calculated in any state $|\psi \rangle$, is greater than or equal to the ground state energy of that Hamiltonian:
\begin{equation}
    \langle \psi | \mathbb{H} |\psi \rangle \geq E_0
\end{equation}

The cost function of the algorithm is straightforward: the expectation value of the problem's Hamiltonian in the parameterized state created by the ansatz
\begin{equation}
    C(\vec{\theta }) = \langle \psi (\vec{\theta })| \mathbb{H} |\psi (\vec{\theta })\rangle
\end{equation}
This is calculated by the sum of the expectation values in the state created by the ansatz for each measurable observable in the Hamiltonian decomposition:
\begin{equation}
    C(\vec{\theta }) = \sum_{i}^{L} h_i \langle \psi (\vec{\theta })| O_i |\psi (\vec{\theta })\rangle
\end{equation}
In the common case of the measurable observables being Pauli operators, measuring their expectation values can be done by adding single qubit gates, before the measurement of the qubits. The Pauli operator can be broken down to a tensor product of single qubit Pauli operators, so the gates to be added for each qubit are simply defined by the local Pauli operator on that qubit. Assuming that the measurement is done in the computational basis, the gates that should be added are as follows - a Hadamard gate for measuring the $X$ operator, a $S^{\dagger}$ gate followed by a Hadamard gate for measuring $Y$ operator. For measuring the $Z$ operator, no gate should be added.

\section{Molecular Electronic Structure Problem}
\label{appendix:D}

One of the most common problems solved using the VQE algorithm is the molecular electronic structure problem. In this problem, an approximation of the ground state energy of the steady state of a molecule should be calculated. The golden standard in computational chemistry for such an approximation is called chemical accuracy and is considered to be the accuracy required to make realistic chemical predictions for some important chemical reactions (for example, the bond distance of a molecule or the energy required to break a molecular bond). This approximation is a fixed (does not scale with problem size) additive approximation equal to 1 [kcal/mol] which is about 1.6 [milli-Hartree/molecule]. Such approximation is made relative to the perfect solution of the problem if one can be calculated analytically, or to the results of a chemical experiment done in a lab in case an analytical solution can not be achieved.

In order to map such problems into a finite space that can be solved on a computer (either classical or quantum), some assumptions and approximations must be used.
In our research we used common assumptions and approximations for the electronic structure Hamiltonian:
\begin{itemize}
\item relativistic effects are completely neglected.
\item The Born–Oppenheimer approximation is used.
\item The molecular orbitals are assumed to be a linear combination of a finite number of basis functions (for example, using the STO-nG family of basis functions, while the $n$ is the number of basis functions).
\item  Only a finite set of molecular spin orbitals are taken into account in the calculation, while the others are defined before the calculation as either empty or fully occupied (the number of spin orbitals will define the problem size and therefore the runtime complexity of the problem). Each molecular orbital is treated as 2 spin orbitals, therefore the number of relevant $M$ spin-orbitals is always an even number.
\end{itemize}

The electronic structure Hamiltonian can be written as follows \cite{jordan_wigner, VQE_best_practices}:
\begin{equation}
\label{FCI_equation}
    \mathbb{H} = \sum_{i,j}^{M} h_{i,j} a_i^{\dagger}a_j + \sum_{i,j,k,m}^{M} h_{i,j,k,m} a_i^{\dagger}a_j^{\dagger}a_ka_m
\end{equation}
while $h_{i,j}$ and $h_{i,j,k,m}$ are the coefficients of the single and two electron energy terms, respectively. When summing over spin orbitals instead of electrons we lose in the equation the conservation of the number of electrons, and might enforce this conversation rule to reduce the number of qubits needed to encode the Hamiltonian. The size of this Hamiltonian (if written as a matrix) is exponential in the number of spin orbitals taken into account in the calculation.
The $h_{i,j}$ and $h_{i,j,k,m}$ terms can be calculated using the one- and two-body integrals:
\begin{equation}
\label{one_body_integral}
\begin{split}
    & h_{i,j} = \frac{1}{2} \langle i| \nabla _i^2 |j\rangle + \langle i| \frac{Z_A}{r_{iA}} |j\rangle = \\
     & \int \phi _i(r)^*(-\frac{\hbar ^2}{2m_e}\nabla ^2 -\sum_{A}^{P} \frac{e^2}{4\pi \epsilon _0}\frac{Z_A}{|r_{i}-R_A|}) \phi _j(r)\,dx
\end{split}
\end{equation}
\begin{equation}
\label{two_body_integral}
\begin{split}
    & h_{i,j,k,m} = \langle i,j| \frac{1}{r_{ij}} |k,m\rangle = \\
    & \frac{e^2}{4\pi \epsilon _0} \int \frac{\phi _i(r)^* \phi _j(r)^* \phi _k(r) \phi _m(r)}{|r_1 - r_2|} \,dx_1\,dx_2
\end{split}
\end{equation}
while $e$ is the elementary charge constant, $m_e$ is the electron mass, $Z_A$ is the atomic number of the nucleus $A$, $R_A$ is the nuclear $A$ position, $r_i$ is the electron $i$ position, $\phi _i$ is the electronic wavefunction of the spin orbital $i$ (represented by a linear set of basis functions) and $x_i$ is subsuming both the electron position and spin ($x_i = (r_i, \sigma)$).

The coefficients of the Hamiltonian are calculated based on the specific molecule of the problem, the basis functions used, the specific orbitals that are taken into account, and the distance (or coordinates in problems with $>1$ dimensions) between the atoms' nucleus. Finding the ground state energy of this Hamiltonian is usually referred as the ``full configuration interaction'' (FCI) problem. The energy of the ground state might be different for each chosen basis functions, orbitals and atoms' coordinates and will be equal or smaller than the Hartree-Fock energy (as additional approximations are taken in the Hartree-Fock energy case).
Solving a similar problem up to a multiplicative error has proven to be a QMA problem \cite{qma}.
There exists a naive method to solve this problem, by diagonalizing the problem's Hamiltonian (as described in equation \ref{FCI_equation}) and finding its smallest eigenvalue. As the size of the Hamiltonian is exponential in the number of spin orbitals taken into account in the calculation, this method is inefficient. One can use this method to compare experimental results to the exact FCI result in small-size problems.

\section{Combinatorial Problems}
\label{appendix:E}

The Max-Cut problem is a well-known combinatorial optimization problem with broad applications in fields such as network design, logistics, and computer science. Given an undirected graph, the goal is to partition the graph's vertices into two disjoint sets (usually denoted as "cutting" the graph) to maximize the total number of edges crossing the partition. This problem is NP-hard, meaning that finding the optimal partition for large graphs becomes computationally challenging.

The cost function we used for this problem is simply summing the number of edges connecting nodes from different subsets (nodes marked as $0$ that are connected to nodes marked as $1$, so the 'cut' passes through them):
\begin{equation}
    C(\vec{x}) = \sum_{i,j \in E(G)} x_i (1-x_j) + (1-x_i) x_j 
\end{equation}
The Max-Cut problem is equivalent to minimizing an Ising model Hamiltonian, as we now show.

In Ising model Hamiltonians defined over a graph $G=(V,E)$, the Hamiltonian is of the form:
\begin{equation}
    \mathbb{H} = -\sum_{i,j\in E(G)}J_{ij}s_is_j
\end{equation}
Here each vertex $i$ of the graph is a spin site that can take a spin value $s_i=\pm 1$. A spin configuration partitions $V(G)$ into two sets, those with spin up $V^+$ and those with spin down $V^-$. We use $\delta (V^+)$ to denote the set of edges that connect the two sets. We can then rewrite the Hamiltonian as:
\begin{equation}
\begin{split}
    \mathbb{H} = -\sum_{i,j\in E(V^+)}J_{ij}-\sum_{i,j\in E(V^-)}J_{ij}+\sum_{i,j\in \delta (V^+)}J_{ij} \\
    = -\sum_{i,j\in E(G)}J_{ij} + 2 \sum_{i,j\in \delta (V^+)}J_{ij} \\
    =C + 2 \sum_{i,j\in \delta (V^+)}J_{ij}
\end{split}
\end{equation}
Where $C$ does not depend on the choice of the cut.
By setting $J_{ij} = -1$, we get that minimizing this Hamiltonian is equivalent to solving Max-Cut for $G$.
Moving from the classical Ising to its quantum counterpart, we get the following equation:
\begin{equation}
\label{eq:ising_maxcut}
    \mathbb{H} = \sum_{i,j \in E(G)}Z_iZ_j
\end{equation}

which can be measured efficiently on a quantum computer by measuring the expectation values of the Pauli $ZZ$ operators.
While Max-Cut is a maximization problem, the best assignment for $G$ would translate to the ground state of $\mathbb{H}$ as defined in Eq.~\ref{eq:ising_maxcut}. Thus, to find the maximum cut for $G$, we seek to minimize the energy of $\mathbb{H}$.



\end{document}